# Simulations for the Extended Hubbard Model Utilizing Nonextensive Statistical Mechanics

F. A. R. Navarro[(1)] and J. F. V. Flores[(2)]

(**1**) farnape@gmail.com, Education National University,

(**2**) jventounmsm@unmsm.edu, National Mayor de San Marcos University

Lima 14, Peru

**Abstract**

We introduce an investigation about M dimers through half-filled two-site Hubbard model, that is, with two electrons. We use the third version of nonextensive statistical mechanics as tool for calculating thermodynamical and magnetic parameters as entropy, internal energy and specific heat. By making computer simulations, we vary the *q* entropic index values between 1 and 2: $q$=1.1, 1.3, 1.5, 1.7 and 2.0. These values are interesting to study magnetic small systems. We find that the addition of intersite interaction term causes a shifting of all simulated curves in relation to simple Hubbard model.

**Key words**: Extended Hubbard model, quantum nonextensive statistical mechanics, and magnetic small systems.

**1. Introduction**

We are motivated for carrying out this work due the interesting results of investigation initiated by H. Hasegawa [1-2], who researched the magnetic small system properties with the simple Hubbard model [3-5]. By using nonextensive statistical mechanics for doing computer simulations with the Newton-Raphson method, He provided evidences for the feasibility of applying the nonextensive theory for magnetic small systems.

Of all the theories known like generalized statistics, the nonextensive statistical mechanics is the more widely studied; it also is known as Tsallis statistics, in tribute to its inventor C. Tsallis. This statistical theory would be able to generalizing the Boltzmann-Gibbs-Shannon statistics [6]. In this paper, we employ the third version of Tsallis statistics proposed in 1998 [7].

In section 2, we display the theoretical frame, the half- filled two-site Hubbard model [8]. Next, we calculate the energy eigenvalues for the Hamiltonian matrix. Those eigenvalues will be used in the thermal average formulas inside the nonextensive statistical mechanics.

In section 3, we introduce the results.

In section 4, we have the conclusions.

## 2. Theoretical Frame

### 2.1 Half-filled Two-site Hubbard Model

The Hubbard model was proposed in 1960's decade by the British physicist John Hubbard [3]. This model is paradigmatic inside the solid state physics; it is a simple model than take into account particles in interaction in a crystal lattice. Through this model, complex phenomena have been explained, namely, metal - insulating transition, ferromagnetic and antiferromagnetic phases and even the superconductivity. In despite of model simplicity, only a few exact solutions for certain cases are known, to see review papers in [3-5]. The problem that we study is N dimers, with 2*N* particles, with N=2M.

The more elementary Hubbard Hamiltonian has two terms: **1)** a kinetic term that allows the electrons jumping between neighbor sites of a crystal lattice, and **2)** a potential energy term that reckons the on-site Colombian interaction. The Hamiltonian is the following one:

$$H_{\text{dimer}} = -t \sum_{\sigma} \left( c^+_{1,\sigma} c_{2,\sigma} + c^+_{2,\sigma} c_{1,\sigma} \right) + U(n_{1,\uparrow} n_{1,\downarrow} + n_{2,\uparrow} n_{2,\downarrow}), \qquad (1)$$

indexes **1** and **2** designate sites of respective dimer, $\sigma$ stand for spines that may be up ($\uparrow$) or down ($\downarrow$). For the kinetic energy term, t is the Hopping integral. Also, inside framework of the second quantization: $c^+_{1,\sigma}$ is the creation operator of a particle in site 1, with spin $\sigma$; $c_{2,\sigma}$ symbolizes the annihilation operator of a particle in site 2, with spin $\sigma$. For the on-site interaction term, *U* is the potential energy, $n_{1,\uparrow}$ represents the particle number operator, in site 1, with spines $\uparrow$ and $n_{1,\downarrow}$ stand for the particle number operator, in site 1, with spines $\downarrow$; the same for site 2.

For this article, we will take into account the extended Hubbard model [8], that is, we add a third energy term, the intersite Coulombian interaction:

$$H_{\text{intersite}} = J_1 \sum_{\sigma} n_{1,\sigma} n_{2,\sigma} + J_2 \sum_{\sigma} n_{1,\sigma} n_{2,-\sigma} \;, \tag{2}$$

$J_1$ and $J_2$ describe interactions between neighbor sites, 1 and 2, inside dimer; they are Coulombians repulsions modified by polaron effects. Consequently, the total Hamiltonian operator that we will study is:

$$\hat{H}_{\text{dimer}} = -t \sum_{\sigma} \left( c^+_{1,\sigma} c_{2,\sigma} + c^+_{2,\sigma} c_{1,\sigma} \right) + U(n_{1,\uparrow} n_{1,\downarrow} + n_{2,\uparrow} n_{2,\downarrow}) + J_1 \sum_{\sigma} n_{1,\sigma} n_{2,\sigma} + J_2 \sum_{\sigma} n_{1,\sigma} n_{2,-\sigma} \;, \tag{3}$$

## 2.1.1 Calculation of Energy Eigenvalues in the Half-filled Two-site Hubbard Model

In this subsection, our goal is to find the energy eigenvalues, so we must make up the Hamiltonian matrix, it is done by using the following basis of six vectors, in bra-ket notation:

$$|\Phi_1\rangle = |\uparrow\downarrow, 0\rangle, \quad |\Phi_2\rangle = |\uparrow, \uparrow\rangle, \quad |\Phi_3\rangle = |\uparrow, \downarrow\rangle, \quad |\Phi_4\rangle = |\downarrow, \uparrow\rangle, \quad |\Phi_5\rangle = |\downarrow, \downarrow\rangle, \quad \text{and} \quad |\Phi_6\rangle = |0, \uparrow\downarrow\rangle, \tag{4}$$

the comma signals separate spines in different sites (if we take the vector basis in a different order then we get another matrix, but the eigenvalues will be the same, for instance, see [9]). Next, the matrix elements of $\hat{H}_{\text{dimer}}$ are built so:

$$[H_{\text{dimer}}]_{m,n} = \langle \Phi_m | \hat{H}_{\text{dimer}} | \Phi_n \rangle \;; \tag{5}$$

in order to obtain the 36 matrix elements, we apply $\hat{H}_{\text{dimer}}$ on respective kets, and afterward, we

obtain:

$$\hat{H}_{dimer}|\Phi_1\rangle = -t(|\Phi_4\rangle+|\Phi_3\rangle)+U|\Phi_1\rangle, \qquad \hat{H}_{dimer}|\Phi_2\rangle = J_1|\Phi_2\rangle,$$
$$\hat{H}_{dimer}|\Phi_3\rangle = -t(|\Phi_1\rangle+|\Phi_6\rangle)+J_2|\Phi_3\rangle, \qquad \hat{H}_{dimer}|\Phi_4\rangle = -t(|\Phi_1\rangle+|\Phi_6\rangle)+J_2|\Phi_4\rangle, \qquad (6)$$
$$\hat{H}_{dimer}|\Phi_5\rangle = J_2|\Phi_5\rangle \qquad \text{and} \qquad \hat{H}_{dimer}|\Phi_6\rangle = -t(|\Phi_3\rangle+|\Phi_4\rangle),$$

then, on the left side from these expressions, we operate with the respective bras and we get the following Hermitian matrix 6x6:

$$H_{dimer} = \begin{pmatrix} U & 0 & -t & -t & 0 & 0 \\ 0 & J_1 & 0 & 0 & 0 & 0 \\ -t & 0 & J_2 & 0 & 0 & -t \\ -t & 0 & 0 & J_2 & 0 & -t \\ 0 & 0 & 0 & 0 & J_1 & 0 \\ 0 & 0 & -t & -t & 0 & U \end{pmatrix}, \qquad (7)$$

and so, we must to diagonalize this matrix to finding the eigenvalues, but it is equivalent to the condition $\det[H_{dimer} - \lambda I] = 0$, $I$ denoting the Identity matrix 6x6. We will apply the properties of determinant for converting it in an upper triangular matrix determinant, and the calculation provides the following expression one:

$$\det[H_{dimer} - \lambda I] = 0 = \begin{vmatrix} U-\lambda & 0 & -t & -t & 0 & 0 \\ 0 & J_1-\lambda & 0 & 0 & 0 & 0 \\ 0 & 0 & J_2-\lambda & \lambda-J_2 & 0 & 0 \\ 0 & 0 & 0 & (U-\lambda)(J_2-\lambda) & 0 & -2(U-\lambda)t \\ 0 & 0 & 0 & 0 & V_1-\lambda & 0 \\ 0 & 0 & 0 & 0 & 0 & \frac{(U-\lambda)^2}{2t}(J_2-\lambda)-2(J_2-\lambda)t \end{vmatrix}; \quad (8)$$

we know that zero is the product of the diagonal elements; straightaway, we perceive that $J_1$ is a two-fold degenerate eigenvalue, $U$, and $J_2$ also are eigenvalues. Next, we build the 2th-degree eigenvalues equation to the two remain eigenvalues:

$$\frac{(U-\lambda)^2}{2t}(J_2-\lambda) - 2(J_2-\lambda)t = 0. \tag{9}$$

Finally, we can establish the six energy eigenvalues (see [08]):

$$\varepsilon_1 = J_2, \quad \varepsilon_2 = U, \quad \varepsilon_3 = C + \frac{U+J_2}{2},$$
$$\varepsilon_4 = -C + \frac{U+J_2}{2}, \quad \varepsilon_5 = J_1 \quad \text{and} \quad \varepsilon_6 = J_1, \tag{10}$$

where,

$$C = \sqrt{\left(\frac{U-J_2}{2}\right)^2 + 4t^2}, \tag{11}$$

### 2.1.1 Calculation of Energy Eigenvectors in Half-filled Two-site Hubbard Model

We display an example of how calculating the energy eigenvectors. We build the matrix expression:

$$\begin{pmatrix} U & 0 & -t & -t & 0 & 0 \\ 0 & J_1 & 0 & 0 & 0 & 0 \\ -t & 0 & J_2 & 0 & 0 & -t \\ -t & 0 & 0 & J_2 & 0 & -t \\ 0 & 0 & 0 & 0 & J_1 & 0 \\ 0 & 0 & -t & -t & 0 & U \end{pmatrix} \begin{pmatrix} c_1 \\ c_2 \\ c_3 \\ c_4 \\ c_5 \\ c_6 \end{pmatrix} = \varepsilon_n \begin{pmatrix} c_1 \\ c_2 \\ c_3 \\ c_4 \\ c_5 \\ c_6 \end{pmatrix}, \tag{12}$$

so for finding the values of constant $c_i$, we build the correspondent linear equations. As an illustrative instance, we get a system of six algebraic equations to $\varepsilon_2 = U$:

$$Uc_1 - tc_3 - tc_4 = Uc_1,$$
$$J_1 c_2 = Uc_2,$$
$$Uc_1 - J_2 c_3 - tc_6 = Uc_3,$$
$$-tc_1 + J_2 c_4 - tc_6 = Uc_4, \qquad (13)$$
$$J_1 c_5 = Uc_5,$$
$$-tc_3 - tc_4 + Uc_6 = Uc_6.$$

We solve it and we find:

$$c_1 = 1, \quad c_2 = 0, \quad c_3 = 0,$$
$$c_4 = 0, \quad c_5 = 0, \text{ y } c_6 = -1, \qquad (14)$$

and so, the eigenvector for $\varepsilon_2 = U$ is get. It can be expressed in column vector or ket notation:

$$X_1 = \begin{pmatrix} 1 \\ 0 \\ 0 \\ 0 \\ 0 \\ -1 \end{pmatrix} = 1 \begin{pmatrix} 1 \\ 0 \\ 0 \\ 0 \\ 0 \\ 0 \end{pmatrix} - 1 \begin{pmatrix} 0 \\ 0 \\ 0 \\ 0 \\ 0 \\ 1 \end{pmatrix} \quad \text{or} \quad |\Psi_1\rangle = |\Phi_1\rangle - |\Phi_6\rangle; \qquad (15)$$

analogously, the other eigenvectors can be obtained.

**2.2 The Third Version of the Nonextensive Statistical Mechanics**

The theoretical construction starts postulating the Tsallis entropy [10]:

$$S_q = k_B \frac{1 - \sum_i (p_i^q)}{q - 1}, \qquad (16)$$

where $p_i^q$ is the probability distribution (strictly, it is the probability density function, PDF) that the system is in *i*-th state, $p_i$, elevated to the entropic index $q$; $k_B$ it is the Boltzmann constant; $\sum_i (p_i^q)$ symbolizes the trace operation over all states of matrix $p_i^q$. In the limit, when $q$ tends to 1, we recover the well known Boltzmann-Gibbs-Shannon entropy:

$$S = -k_B \sum_i (p_i \mathrm{Ln} p_i), \tag{17}$$

The probability distribution $p_i$ is obtained through maximum entropy method that was invented by American Edward T. Jaynes [11]. In that procedure, we consider the following constraints ones:

$$\sum_i p_i = 1 \quad \text{and} \quad U_q = \frac{\sum_i p_i^q \varepsilon_i}{\sum_i p_i^q}, \tag{18}$$

$\varepsilon_i$ are the energy eigenvalues. With the maximum entropy method, we obtain the probability distribution:

$$p_i = \frac{[1-(1-q)\beta'\varepsilon_i]^{\frac{1}{1-q}}}{Z_q}, \tag{19}$$

the expression $[1-(1-q)\beta'\varepsilon_i]^{\frac{1}{1-q}}$ is the known $q$-exponential function, and $Z_q$ is the partition function given by:

$$Z_q = \sum_i [1-(1-q)\beta'\varepsilon_i]^{\frac{1}{1-q}}, \tag{20}$$

with $\beta'$ an energy parameter:

$$\beta' = \frac{1}{k_B T}, \tag{21}$$

*T* is the temperature of the system. It should be said that some authors use other temperature concepts, for it is an open problem [7, 12].

We want emphasize that in the Eq. (19) is necessary impose the Tsallis cut-off:

$$1-(1-q)\beta'\varepsilon_i \geq 0, \tag{22}$$

so that the definition to distribution of probability is:

$$p_i = \begin{cases} \dfrac{[1-(1-q)\beta'\varepsilon_i]^{\frac{1}{1-q}}}{Z_q}, & \text{if } 1-(1-q)\beta'\varepsilon_i \geq 0 \\ 0, & \text{otherwise} \end{cases} \qquad (23)$$

**2.1 Quantum Average Values**

In the third version of the nonextensive statistical mechanics, average values to any observable $\hat{O}$ in the Hilbert space are calculated by means of the formula:

$$\langle \hat{O} \rangle = \frac{\sum_i p_i^q O_i}{\sum_i p_i^q}, \qquad (24)$$

$O_i$ are the eigenvalues of the observable $\hat{O}$.

For instance, we have that the internal energy is given by:

$$E_{\text{internal}} = \langle \hat{H} \rangle = \frac{\sum_i p_i^q \varepsilon_i}{\sum_i p_i^q}, \qquad (25)$$

with this parameter we can get the specific heat:

$$C_e = \frac{\partial E_{\text{internal}}}{\partial T}, \qquad (26)$$

**2.1 Alternative Derivation to $q$-Exponential Function**

The idea is very simple; we start with the definition of Euler's Number:

$$e = \lim_{n\to\infty}\left(1+\frac{1}{n}\right)^n = \lim_{h\to 0}(1+h)^{\frac{1}{h}}, \qquad (27)$$

we can broaden this definition for the exponential function that can be expressed in several ways:

$$e^x = \lim_{h\to 0}(1+hx)^{\frac{1}{h}} = \lim_{z\to 1}[1+(z-1)x]^{\frac{1}{z-1}} = \lim_{q\to 1}[1+(1-q)x]^{\frac{1}{1-q}}, \qquad (28)$$

in the limit operation, we made the respective change of variables. Essentially, in Eq. (28), the last expression is the same as Eq. (19) that is get by maximum entropy method.

## 3. Computer simulations

We utilize the programming language Matlab 7.0 for doing computer simulations of the following thermodynamic properties: entropy per dimer, internal energy per dimer and specific heat per dimer. For simplifying the simulations, we define the new variables $T/t$, $U/t$, $E/t$ and $J/t$; additionally, we assume that $J_1 = J_2$ and $k_B = 1$, so that the entropy is in Boltzmann constant units.

The figure 1 shows the entropy vs. the normalized temperature, that is, $S_q$ vs. $T/t$, with entropic index, $q$=1.1, 1.3, 1.5, 1.7 and 2.0, added parameters are to **a)** $U/t = 6$ and $J/t=0$; to **b)** $U/t =12$ and $J/t=0.1$. For both graphics, to low temperatures, it is seen that entropy is zero for all values of $q$. The more enlarge the value of q the more diminish saturation entropy. In the entropy curves, the addition of $J/t$ causes a small shifting on the right side. This is due to the fact that $J/t$ modifies the condition for Tsallis cut-off, that is, Eqs. (22) and (23).

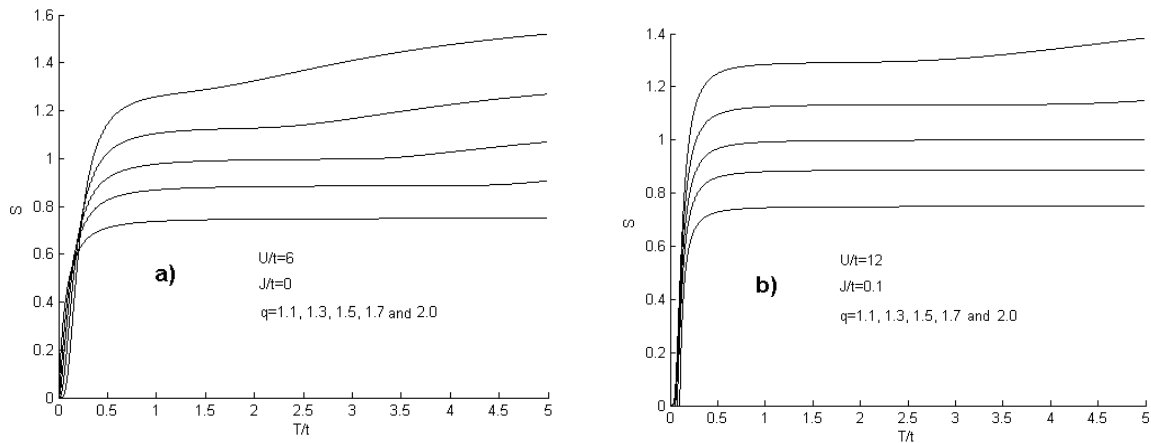

**Fig 1.** Entropy vs. normalized temperature. The values of parameters are indicated inside each individual graph.

The figure 2 shows the normalized internal energy vs. the normalized temperature, that is, *U/t vs. T/t*, with entropic index, *q*=1.1, 1.3, 1.5, 1.7 and 2.0, added parameters are for **a)** *U/t* = 6 and *J/t*=0; for **b)** *U/t* = 12 and *J/t*=0.1. For both graphics, the more increase *U/t* the more increase internal energy, it is due positive term of Coulombian repulsion in Eq. (3). The addition of intersite interaction term increase more the positive shiftings.

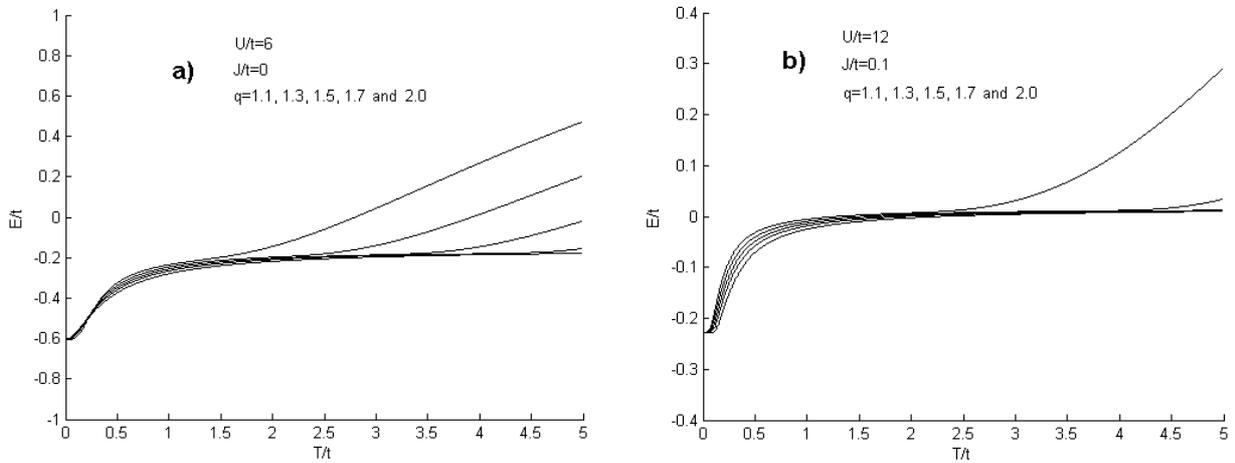

**Fig 2.** Internal energy *vs.* normalized temperature. The values of parameters are indicated inside each individual graph.

The figure 3 displays graphs of the specific heat *vs.* the normalized temperature, *Ce* vs. *T/t*, with entropic index, *q*=1.1, 1.3, 1.5, 1.7 and 2.0, added parameters are for **a)** *U/t* = 6 and *J/t*=0; for **b)** *U/t* = 12 and *J/t*=0.1. For both graphics, the more increases the value of *q* the more diminish the peak heights. Also, we observe that some valleys are formed for each value of *q*; the more increase q, the more diminish depth of valleys. It practically disappearing in *q*=2.0. We also observe small shifting on the right side.

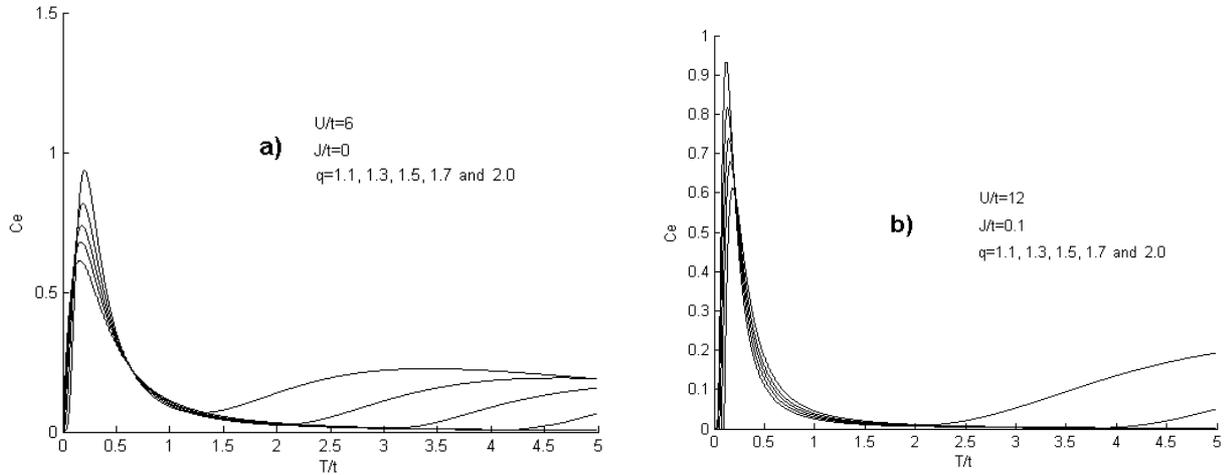

**Fig 3.** Specific heat vs. normalized temperature. The values of parameters are indicated inside each individual graph.

## 4. Conclusions

In this article, we analyzed *N* dimers with a half-filled two-site Hubbard model. Together with this model we utilize the third version of nonextensive statistical mechanics as tool for calculating entropy, internal energy and specific heat magnetic. The computer simulations for Hubbard model with on-site term are in total agreement with previous results. The addition of intersite interaction term produces a shift in all the parameters curves. This is due to intersite interaction term modifies the condition for the Tsallis cut-off.